\documentclass[12pt]{article}
\usepackage{epsf}
\font\af=msbm12
\font\aff=eurm10
\newcommand{\eqnn}[1]{\begin{eqnarray*}#1\end{eqnarray*}}
\newcommand{\eqnl}[2]{\par\parbox{14.5cm}
{\begin{eqnarray*}#1\end{eqnarray*}}\hfill
\parbox{1cm}{\begin{eqnarray}\label{#2}\end{eqnarray}}}

\newcommand{\eqngrlb}[3]{\par\parbox{12.5cm}
{\begin{eqnarray}\fbox{$\displaystyle
#1\\#2$}\end{eqnarray}}\hfill
\parbox{1cm}{\begin{eqnarray}\label{#3}\end{\eqnarray}}}

\newcommand{\refs}[1]{(\ref{#1})}

\def\om{\omega}
\def\ms{\!-\!}
\def\vx{\vec{x}}
\def\intl{\int\limits}
\def\di{\displaystyle}
\def\lam{\lambda}
\def\&{&\di}
\def\bg{\begin{eqnarray}\begin{array}{rcl}\displaystyle}
\def\eg{\end{array} &\di    &\di   \end{eqnarray}}
\def\bm#1{\begin{eqnarray}\begin{array}{#1}\di} 
\def\bmo#1{\begin{eqnarray*}\begin{array}{#1}\di} 
\def\eg{\end{array} &\di    &\di   \end{eqnarray}}
\def\bgo{\begin{eqnarray*}\begin{array}{rcl}\displaystyle}
\def\ego{\end{array} &\di    &\di \nonumber  \end{eqnarray*}}

\def\btensor#1#2{\renew\left#1\begin{array}{#2}\di}
\def\etensor#1{\end{array}\right#1}
\def\alzc{\alpha^\vee_{(0)}}
\def\alz{\alpha_{(0)}}
\def\muzc{\mu^\vee_{(0)}}

\def\ha{{1\over 2}}

\def\d{{\mbox d}}

\def\Tr{\mbox{Tr}}

\def\T{{\mbox T}}

\def\id{1\!\mbox{l}}

\def\ov{\over}
\def\qinst{q}

\def\vex0{\vx _{\rm d}}

\def\al{\alpha}
\def\pr{\prime}

\def\eps{\epsilon}

\def\R{\mbox{\af R}}

\def\T{\mbox{\af T}}
\def\w{\mbox{\aff w}}
\def\Z{\mbox{\af Z}}
\def\CA{{\cal A}}

\def\CC{{\cal C}}
\def\CD{{\cal D}}

\def\CF{{\cal F}}

\def\CH{{\cal H}}

\def\CM{{\cal M}}

\def\CP{{\cal P}}
\def\CS{{\cal S}}

\def\CZ{{\cal Z}}
\def\CW{{\cal W}}

\def\pan{\par\noindent}

\def\musc{\mu^\vee_{(\sigma)}}
\def\als{\al_{(\sigma)}}
\def\alsc{\al^\vee_{(\sigma)}}
\newcommand{\mtxt}[1]{\quad\hbox{{#1}}\quad}

\date{\today}

\newcommand{\mysection}[1]{\section{#1}\setcounter{equation}{0}}

\def\renew{\renewcommand{\arraystretch}{1}}

\begin{document}

\large \centerline{Abelian Projection on the Torus for general
Gauge Groups\footnote{Supported 
by the Deutsche Forschungsgemeinschaft,
DFG-Wi 777/3-2}} 
\vskip 1truecm
\normalsize

\begin{center}
\textbf{C.~Ford, T.~Tok and A.~Wipf}\footnote{e--mails:
Ford, Tok and Wipf@tpi.uni-jena.de}\\
\it{Theor.--Phys. Institut, Universit\"at Jena\\ 
Fr\"obelstieg 1, D--07743 Jena, Germany}
\end{center}
\par
\vskip 1 truecm
\begin{abstract} 
We consider Yang-Mills theories with general gauge groups $G$ 
and twists on the four torus. We find consistent boundary conditions 
for gauge fields in all instanton sectors.
An extended Abelian projection with respect
to the Polyakov loop operator is presented, where $A_0$ is independent 
of time and in the Cartan subalgebra. Fundamental domains for the
gauge fixed $A_0$ are constructed for arbitrary gauge groups. 
In the sectors with non-vanishing instanton number such gauge fixings
are necessarily singular. The singularities can be restricted to Dirac
strings joining magnetically charged defects.
The magnetic charges of these monopoles take their values
in the co-root lattice of the gauge group.
We relate the magnetic charges of the defects and the
windings of suitable Higgs fields about these defects
to the instanton number.
\end{abstract}
      

\noindent PACS numbers: 11.10Wx, 11.15Tk, 11.15Kc, 12.38Aw 

\vskip .5 truecm

\noindent {\bf Keywords:} Gauge field theory at finite temperature, 
gauge fixing, abelian projection, magnetic monopoles, instanton number

\mysection{Introduction}

Confinement and chiral symmetry breaking are
supposed to follow from the dynamics of Yang-Mills fields.
These phenomena are highly non-perturbative and still
have not been derived from first principles.
In this paper we will follow the strategy put forward by 't Hooft
\cite{tH} who considered Yang-Mills theories on a
Euclidean space-time torus $\T^4$. The torus provides a
gauge invariant infrared cut-off. Its
non-trivial topology gives rise to a non-trivial structure
in the space of Yang-Mills fields which yields additional
information on the possible phases of Yang-Mills theories.
Compared to other Riemannian 4-dimensional compact manifolds
the torus has many advantages (besides being the `space-time'
used in lattice simulations):
\begin{itemize}
\item
one can use a flat metric in which case
curvature effect do not mix with finite size effects,
\item
the circumference $L_0$ in the temporal direction can
be identified with the inverse temperature $\beta$ \cite{pisarski,kapusta},
\item
gauge invariant periodic fields on $\R^4$ can be
viewed as fields on $\T^4$,
\item
one may calculate non-perturbative quantities
from finite size effects \cite{luesch}; 
the string constant is directly related to the energy of a string
winding around the torus \cite{tH},
\item
one keeps the relevant part of the supersymmetry in SUSY-YM theories.
\end{itemize}
Even the less ambitious goal to demonstrate confinement 
of static quarks without reliance on numerical simulations
has not been achieved yet. Without dynamical fermions the
relevant observables are products of Wilson-loops \cite{wilson}. 
At finite temperature $T=1/\beta$ the gauge fields
in the functional integrals
are periodic in Euclidean time i.e.
\eqnn{
A_\mu(x^0+\beta,\vx)=A_\mu(x^0,\vx).}
and one may use Polyakov loops \cite{polloops}
\eqnl{
P(\vx)=\Tr~ R\big(
\CP (\beta,\vx)\big),
\mtxt{where}\CP (x^0,\vx)=\CP \exp\left[
i\int^{x^0}_0 d\tau A_0(\tau,\vx)\right]}{defpol}
as order parameters for confinement. Here
$R$ is the representation of the gauge group which acts on
the matter fields. We shall assume 
that the \textit{gauge group $G$ is simply connected}, e.g. $G=SU(2)$ 
rather than $SO(3)=\rm{Ad}(SU(2))$. But since we allow
for arbitrary representations $R$ of $G$ our results apply
to general gauge groups $R(G)$, for
example to $SO(3)$.

The Polyakov loop $P(\vx)$ is invariant under 
gauge transformations which are periodic in time.
Since it is a functional of $A_0$ only,
one is motivated to seek a gauge fixing where $A_0$ is as simple as 
possible. Note  that the Weyl gauge, $A_0=0$, is not compatible 
with time-periodicity. 
In a previous paper \cite{us} we discussed an extended
Abelian projection for $SU(2)$ gauge theories on the four torus
in which $A_0$ is time independent and in the Cartan subalgebra.
The gauge fixing procedure hinges on the  diagonalization
of the path ordered exponential, $\CP(\beta,\vx)$,
whose trace is the Polyakov loop.
In contrast to the two dimensional case investigated in \cite{mpw} the 
diagonalization procedure has unavoidable  singularities
\cite{tHooft,schierholz}. The
singularities can be interpreted as Dirac strings \cite{dirac} joining 
magnetically charged  `defects'. 
Here we understand defects as points, loops  (not to be confused with
the Dirac strings!), sheets and lumps where $\CP(\beta,\vx)$
has degenerate eigenvalues.
For the gauge group $SU(2)$, the eigenvalues of
$\CP(\beta,\vx)$  are degenerate when $\CP(\beta,\vx)=\pm \id$.
Thus there are two types of defect according to whether
$\CP(\beta,\vx)$ is plus or minus the identity.
Associated with the gauge fixing procedure
one can define an Abelian magnetic potential $A_{mag}$ 
on $\T^3$ \cite{tHooft}.
In \cite{us} we showed that the total magnetic charge of
$\CP=\id$ defects is equal to the instanton number
$\qinst$. 
Moreover, the total magnetic charge of all defects is zero, i.e.
the total magnetic charge of 
$\CP=-\id$ defects is minus that of the $\CP=\id$ defects.
The relationship between magnetic charges and the instanton number
was considered earlier by Christ and Jackiw \cite{Jackiw}, Gross et.al.
\cite{pisarski} and Reinhardt \cite{reinhardt} who
worked on $ S^1 \times \R^3$ or $\R^4$. 
Though here one requires `charges at infinity'
to have overall magnetic charge neutrality.
For an explicit discussion of the singularities emerging in the 
gauge fixing procedure at point like monopoles see the recent 
paper by Jahn and Lenz \cite{Lenz3}.

In this paper we extend the defect analysis to
gauge theories on $\T^4$ with arbitrary gauge groups $G$ of
rank $r$.
We also consider arbitrary twists \cite{tH}, 
which allows us to treat matter transforming
according to any representation of the gauge group.
One has $r+1$ types of \textit{basic defects} associated with the $r+1$
faces constituting the boundary of a `fundamental domain'
(these are essentially compactified Weyl chambers)
in the root space.
Since the magnetic potential lies in the Cartan subalgebra
$\CH$ 
we now have a matrix $Q_M\in\CH$ of magnetic charges. The
possible magnetic charges are quantized and are 
in one to one correspondence
with the points of the integral co-root lattice.
For a basic defect, $Q_M$ is
an integer multiple of a fixed matrix.
Much as in the $SU(2)$ analysis there is a simple linear relation
between the total magnetic charge
of a given type of defect and the instanton number $\qinst$.
We have overall charge neutrality on $\T^3$ 
unless there are non-orthogonal magnetic and electric twists.

The paper is organized as follows.
In the remainder of this section we recall some basic facts concerning
gauge fields on $\T^4$.
Next we present a set of transition functions (i.e. boundary
conditions for the gauge fields) where the instanton
number is equal to the winding number of the mapping 
${\cal P}(\beta,\vx)
:\T^3\rightarrow G$. These transition functions serve 
as the starting point for our gauge fixing.
In section three we construct `fundamental domains' for all gauge groups.
Our Lie algebra conventions are stated here.
Then we explain precisely what we mean by `defects'.
In the next section we define the magnetic charge of the defects.
Our key result is given in section six. Here we obtain the relationship
between the magnetic charges and the instanton number.
Next we rewrite ${\cal P}(\beta,\vx)$ in terms of `Higgs fields'.
This enables us to tie up a loose end from section six,
and also allows us to interpret the magnetic charges
as Higgs winding numbers.
In section eight
we show how the ideas apply to $SU(3)$ and give our conclusions
in section nine.
Technicalities regarding our transition functions 
(including a construction of magnetic twist eaters for all
gauge groups)
can be found in Appendix A.
Finally, an identity quoted in section six is derived in
Appendix B.

We view the four torus as $\R^4$ modulo the lattice generated by
four orthogonal vectors $b_\mu,\;\mu=0,1,2,3$, for a recent
review see \cite{arroyo}.
The Euclidean lengths of the $b_\mu$ are denoted by
$L_\mu$ (we may identify $L_0$ with the inverse temperature $\beta$).
Local gauge invariants such as $\hbox{Tr}\,F_{\mu \nu} F_{\mu \nu}$
are periodic with respect to a shift by an arbitrary lattice vector. 
However, the gauge fields have to be periodic only up to gauge 
transformations. In order to specify boundary conditions for gauge potentials
$A_\al$ on the torus one requires a set of \textit{group valued transition
functions }
$U_\mu(x)$, which are defined on the whole of $\R^4$.
The periodicity properties of $A_\al$ are as follows
\eqnn{
A_\al(x+b_\mu)
=U_\mu^{-1}(x)A_\al(x)U_\mu(x)+i
U_\mu^{-1}(x)\partial_\al U_\mu(x),
\quad \al,\mu=0,1,2,3,}
where the summation convention is \textit{not} applied.
It follows at once, that the path ordered exponential
$\CP (x^0,\vx)$ in \refs{defpol} has the
following periodicity properties 
\eqnl{
\CP(x^0\!+\!L_0,\vx )=\CP(x^0,\vx )\CP (L_0,\vx),\quad\quad
\CP (x^0,\vx \!+b_i)=U_i^{-1}(x^0,\vx )\CP (x^0,\vx )
U_i(0,\vx).}{polloopperiod}
The transition functions 
$U_\mu(x)$ satisfy the  cocycle conditions \cite{tH}
\eqnl{
U_{\mu}(x)U_{\nu}(x+b_\mu)=z_{\mu\nu}U_\nu(x)U_\mu(x+b_\nu),\qquad
z_{\mu\nu}=z^{-1}_{\nu\mu},}{cocycle}
where the \sl twists \rm $z_{\mu\nu}$ lie in the center $ {\cal{Z}} $ 
of the group.
From now on \sl we assume that the transition functions 
belong to the universal covering group. \rm
In general, our matter fields will not transform according
to the covering group.
However, a matter field in some representation is equivalent
to matter transforming according to the covering group
{\it provided we place suitable restrictions on the twists}. \rm
More precisely, consider a matter field which transforms
under some representation $R(G)$ of the gauge group.
A  center element $z\in {\cal{Z}}$ is an allowed twist
if $R(z)=\id$.
For example if we have matter fields in the defining
representation of $SU(3)$ all the twists must be the identity, since
the other two center elements are faithfully represented.
By contrast, if the matter fields are in the adjoint representation of 
any group then there is no restriction on the twists.

Under a gauge transformation, $V(x)$, the pair $(A,U)$ is mapped to 
\eqnl{ 
A_\al^V(x)=V^{-1}(x)A_\al(x)V(x)+i
V^{-1}(x)\partial_\al V(x),\quad
U_\mu^V(x)=V^{-1}(x)U_\mu(x)V(x\!+b_\mu).}{newU}
The twists, $z_{\mu\nu}$, are gauge invariant.
We define the topological charge or instanton number as follows
\eqnl{
\qinst={1 \ov 32\pi^2}
\intl_{\T^4}\eps_{\mu\nu\al\beta}\,\hbox{Tr}\,F_{\mu\nu} 
F_{\al\beta},}{topological}
where the trace corresponds to the canonically normalized scalar product 
in the Lie algebra\footnote{It is equal to half the trace in the adjoint
representation divided by the dual Coxeter number.}. 
Note that $\qinst$
is fully determined by the transition functions \cite{vanbaal}.
In particular, if we take all the transition functions to be the
identity (i.e. we assume the gauge fields are periodic in all directions)
then the instanton number is zero. Accordingly, if we are to 
describe the non-perturbative  sectors, one must consider 
non-trivial transition functions.
For a given $\qinst$ and set of twists, $z_{\mu\nu}$,
we only require \textit{ one } set of transition functions.
If we have two sets of transition functions with the same
instanton  number and twists  then they are gauge equivalent \cite{vanbaal}.

\mysection{Transition functions, the Polyakov loop operator and gauge fixing}

First we construct a convenient set of transition function such that
the instanton number is equal to the winding number of the map
$\CP(\beta,\vx):\T^3\to G$. Then we find the (in general singular)
gauge transformation which transforms $A_0$ into a time-independent
field in the Cartan subalgebra.\pan
In the untwisted case, $z_{\mu\nu}=\id$, we may assume that the transition
functions have the following properties
\eqnl{
U_0=\id,\ \ U_i(x^0\!=\!0,\vx )=\id,\qquad i=1,2,3,\mtxt{so that}
U_i(x+b_0)=U_i(x).}{condition}
In \cite{us} it was shown by explicit construction that there exist untwisted
(i.e. $z_{\mu \nu} = \id$) 
transition functions satisfying \refs{condition} in all
instanton sectors.
The condition that $U_0=\id$ is simply the statement that our
gauge fields are periodic in time.
Since the transition functions are trivial on the time slice
$x^0=0$, and hence with \refs{condition} also on the time slice
$x^0=\beta$, the path ordered exponential
$\CP(\beta,\vx )$ is periodic in the three spatial directions
(see \refs{polloopperiod}).

In the presence of magnetic twists (i.e. at least one of the
$z_{ij}\neq\id$) it is no longer possible
to attain (\ref{condition}).
However, one can still arrange for the transition functions
to be independent of $\vx $ on the time slice
$x^0=0$.
In appendix A we prove that there exist transition functions 
with the following
properties
\eqnl{
U_0=\id,\quad
U_i(x^0=0,\vx)=\om_i,\quad\mtxt{so that}
U_i(x^0=\beta,\vx)=\om_i z_{0i},}{twistcondition}
where the $\om_i$ are independent of $\vx $ and satisfy the
`twist eating' conditions
\eqnl{
\om_i\om_j=z _{ij}\om_j\om_i,\quad
i,j=1,2,3,}{twisteaters}
which follow from the cocycle conditions for the $U_i$ at time
$x^0=0$. For example,
consider $SU(2)$ gauge theory with the following magnetic
twists $z_{12}=-\id, \,\,\, z_{23}=z_{31}=\id$.
Then a possible choice of $\om_i$'s
is $\om_1=i\sigma_1$,
$\om_2=i\sigma_2,\;\om_3=\id$, where the $\sigma_i$
are the Pauli matrices.
Twist eaters satisfying (\ref{twisteaters}) are known to exist
for arbitrary twists in $SU(N)$ gauge theories
\cite{ambjorn}. Twist eaters for the other simple Lie groups
are constructed in appendix A.

Now we use the properties of the transition functions to obtain
a relation for the instanton number in terms of the Polyakov loop.
Consider the following gauge transformation
\eqnn{
V(x^0,\vx )=\CP (x^0,\vx ),}
where $\CP (x^0,\vx )$ is the path ordered exponential
in \refs{defpol} which in general is non-periodic in time.
For brevity we use the notation
\eqnl{\CP(\vx):=\CP(\beta,\vx).}{abbreviation}
Using (\ref{polloopperiod},\ref{newU},\ref{condition}), the gauge 
transformed transition functions are
\eqnn{
U_0^V=\CP(\vx ),\qquad U_i^V=\om_i.}
The new $U_0$ is simply the path ordered exponential 
$\CP(\vx)$, while the transformed spatial transition functions
are constant matrices. Applying the well know formula
for the instanton number in terms of the transition functions 
\cite{vanbaal} yields
\eqnl{
\qinst=\frac{1}{24\pi^2}\intl_{\T^3}\eps_{0 i j k}\hbox{Tr}\left[
(\CP^{-1}\partial_i \CP)(\CP^{-1}\partial_j \CP)(\CP^{-1}\partial_k
\CP)\right],}{polloopindex}
where $\CP=\CP(\vx )$, and $\T^3=\{x\in \T^4|x^0=0\}$.
We emphasize that \refs{polloopindex} is only valid
when the (original) transition function satisfy \refs{twistcondition}.
Another useful consequence of (\ref{twistcondition})
is that $\CP({\vx})$ has very simple periodicity
properties
\bg
\CP(\vx+b_i)=z_{0i}\,\om_i^{-1}\CP(\vx)\,\om_i,\quad i=1,2,3.\eg
In particular, $\CP(\vx)$ is completely periodic in the absence of twists.

Now we follow \cite{Weiss,Langmann,Lenz2,us,mpw} and seek a (time-periodic) 
gauge transformation, $V(x)$, for which the gauge transformed 
$A_0$ is independent of time and in the Cartan subalgebra.
Consider the time-periodic gauge transformation
\eqnl{
V(x^0,\vx )=\CP(x^0,\vx )\, {\CP}^{- x^0/\beta}( \vx)\,W(\vx ),}
{gauge-trf-1}
where $\CP(x^0,\vx )$ is the path ordered exponential \refs{defpol},
and $W(\vx)$ diagonalizes $\CP(\vx)$, i.e.
\eqnl{
\CP(\vx)=W(\vx) D(\vx) W^{-1}(\vx ),\qquad
D(\vx)=\exp [ 2\pi i\, h(\vx ) ],}{diagonalization}
with $h(\vx )$ in the Cartan subalgebra ${\cal H}$.
The fractional power of $\CP$ is defined via the
diagonalization of $\CP$.
It follows at once that the gauge transformed $A_0$ reads
\bg
A_0^V=\frac{2\pi}{\beta}h(\vx),\eg
which is indeed independent of time and in the Cartan subalgebra.
Whereas $\CP(\vx)$ is smooth the factors $W(\vx)$ and
$D(\vx)$ in the decomposition \refs{diagonalization} are in
general not. The classification and implications of these
singularities are investigated in sections 4-7.
\def\ali{{\al_{(i)}}}
\def\alj{{\al_{(j)}}}
\def\mui{{\mu_{(i)}}}
\def\muj{{\mu_{(j)}}}
\def\alic{\al^\vee_{(i)}}
\def\aljc{\al^\vee_{(j)}}
\def\muc{\mu^\vee}
\def\muic{\mu^\vee_{(i)}}
\def\mujc{\mu^\vee_{(j)}}
\def\smui{\vert\mui\rangle}
\def\smuj{\vert\muj\rangle}
\def\Hali{H_{\ali}}

\mysection{Fundamental domains}
\label{sec-Lie-conv}

The mapping $h(\vx)\to D(\vx)$ in \refs{diagonalization} from the
Cartan subalgebra to the toroidal (Cartan) 
subgroup is not one to
one. In this section we shall find domains $\CM$ in 
the Cartan subalgebra such that this mapping
becomes bijective. We shall choose domains which are
left invariant under the action of the Weyl group $\CW$. 
If $\w$ is a Weyl reflection, then 
$W\w$ diagonalizes $\CP$ in \refs{diagonalization} if $W$ does.
We shall fix this residual gauge freedom, under which
$D\to \w D\w^{-1}$, by restricting 
$h$ to one Weyl chamber. The intersection of a Weyl chamber
with the `Weyl invariant' domain $\CM$
defines our fundamental domain $\CF$.
$\CF$
 is  in one to one 
correspondence with the toroidal subgroup modulo Weyl transformations
or equivalently with the conjugacy classes of $G$.
The main result of this section is that $\CF$ is the 
simplicial box with the extremal points \refs{wc4}.

Our Lie algebra conventions are as follows:
Let $H_k,\;k=1,\dots,r$ be an
orthogonal basis of the Cartan subalgebra $\CH$,
$$
\Tr \,H_k H_l = {\vert\al_L\vert^2\ov 2}\delta_{k l},
$$
which are diagonal in a given representation\footnote{We use
the same symbol $H_k$ for $H_k$ in any representation.},
\eqnn{
H_k\vert\mu\rangle=\mu_k\vert\mu\rangle\mtxt{and}
[H_k,E_\al]=\al_k E_\al.}
We normalize the roots such that the long roots
have length $\sqrt{2}$, i.e. $(\al_L,\al_L)=2$,
and the $H_k$ become orthonormal.
{\it Throughout this paper we identify 
$\sum \rho^k H_k=\rho\cdot H \in \CH $ with
$\rho\in \R^r$.} \rm Let
\eqnl{
\ali\,\,,\quad\mui\,\,,\quad
\alic={2\ali\ov \big(\ali,\ali\big)}
\mtxt{and}\muic={2\mu_{(i)}\ov \big(\ali,\ali\big)}\,\,,
\qquad i=1,\dots,r}{sfc}
be the simple roots, fundamental weights, co-roots and
co-weights, respectively:
\eqnl{
\big(\ali,\aljc\big)=K_{ij},\quad
\big(\alic,\muj\big)=
\big(\ali,\mujc\big)=\delta_{ij},\quad
\big(\mui,\mujc\big)=(K^{-1})_{ij} . }{gf1}
We used that the simple roots and fundamental 
weights are related by the Cartan matrix,
\eqnn{
\ali=\sum_{j=1}^r K_{ij}\;\muj.}
The fundamental weight-states (which are the highest weight
states of the $r$ fundamental representations) and
states in the adjoint representation obey
\eqnl{
\alic\cdot H\;\vert\muj\rangle=
\delta_{ij}\vert\muj\rangle\mtxt{and}
\muic\cdot H\;\vert\alj\rangle=\delta_{ij}\vert\alj\rangle.}{reps}
The most negative root $\alz$ and its co-root $\alzc$ define the
integral \textit{Coxeter numbers} $n_i$ and \textit{dual Coxeter numbers}
$n^\vee_i$:
\eqnn{
0=\alz+\sum_1^r n_i\ali\equiv \sum_{\sigma=0}^r n_\sigma\als\mtxt{and}
0=\alzc+\sum_1^r n^\vee_i\alic\equiv \sum_{\sigma=0}^r n^\vee_\sigma \alsc,}
where we have defined $n_0=n^\vee_0=1$. The (dual) Coxeter numbers are listed
in appendix A. For later convenience we assign to $\alz$ 
the co-weight $\muzc=0$.

The fundamental domains we seek
are intimately related to the center elements of the group.
Thus it is useful to find conditions on $\rho\cdot H\in \CH$ 
such that $\exp(2\pi i\rho\cdot H)$ is in the center $\CZ$. 
Center elements are the identity in the adjoint
representation. Because of the second set
of equations in \refs{reps} they must be powers of
\eqnn{
z_i=\exp\Big(2\pi i \muic\cdot H\Big).}
In an irreducible representation a center element acts the same
way on all states. Hence, a necessary and sufficient condition
for $z_i\neq \id$ is that
\eqnn{
z_i\,\vert\muj\rangle=\exp\Big(2\pi i K^{-1}_{ji}\Big)\vert\muj\rangle
\neq \vert\muj\rangle,\mtxt{or that}  K^{-1}_{ji}\notin \Z}
for at least one fundamental weight $\muj$. Here we have used that
the inner products of the weights with the co-weights yield the inverse
Cartan matrix, see \refs{gf1}.
The order of the center group is just $\det(K)$.
The centers and their generators are  listed
in appendix A.
\pan
Let us now find a suitable domain in the Cartan subalgebra which
is mapped bijectively into the toroidal subgroup. The elements
\eqnn{
\exp\Big(2\pi i\rho\cdot H \Big)}
in the toroidal subgroup are the identity if 
$\rho$ is in the integral co-root lattice, i.e. the lattice spanned 
by the simple co-roots $\al^\vee_{(i)}$ (see \refs{gf1}). 
Thus, the convex region $\CM$ defined by
the intersecting half-spaces $(\rho,\al)\leq 1$, where $\al$ is
an arbitrary root, is in one to one\footnote{On the boundary of the so defined
set we have to identify points differing by a vector $\al^\vee$, i.e. we have
to remove half of the boundary to get a one to one correspondence.}
correspondence with the 
toroidal subgroup of the gauge group\footnote{The hyperplane
$(\rho,\al)=1$ is orthogonal to $\al^\vee$ and goes
through $\al^\vee/2$.}. This set is invariant under
the action of the Weyl group $\CW$ and is given by
\eqnl{
\CM=\{\rho\vert\;\; (\rho,\al)\leq  1\mtxt{for all roots }\al\}.}{wc1}
Now we may fix the residual Weyl reflections
by further assuming that $\rho\sim \rho\cdot H$ is in the Weyl chamber
defined by
\eqnl{
\{\rho\vert\;\; (\rho,\ali)\geq 0\quad\mtxt{for 
all simple roots}\ali\}.}{wc2}
The inner product of a vector $\rho$ in this Weyl chamber
with the highest root $-\alz$ is always greater or equal to
the inner product with any other root. It follows
that the conditions (\ref{wc1},\ref{wc2}),
which define the \textit{fundamental domain} $\CF$,
simplify to 
\eqnl{
\CF=\Big\{\rho \vert \;\;(\rho,\ali)\geq 0,\quad -(\rho,\alz)
\leq 1\Big\}.}{weylch}
$\CF$ is a simplex bounded by $r+1$ hyperplanes
orthogonal to the roots $\{\als\}=\{\alz,\ali\}$. 
In what follows we call the plane orthogonal to $\als$
the $\sigma$-plane, $\sigma\in\{0,i\}$. The 
$i$-planes all meet at the origin.
Since $\alz$ is a long root
the last condition in \refs{weylch} means that the $0$-plane
orthogonal to $\alz$ goes through $-\alzc/2$.
The roots $\als$ point inside the box.

\begin{figure}[ht]
\begin{minipage}[ht]{16cm}
\centerline{\epsfysize=13 cm\epsffile{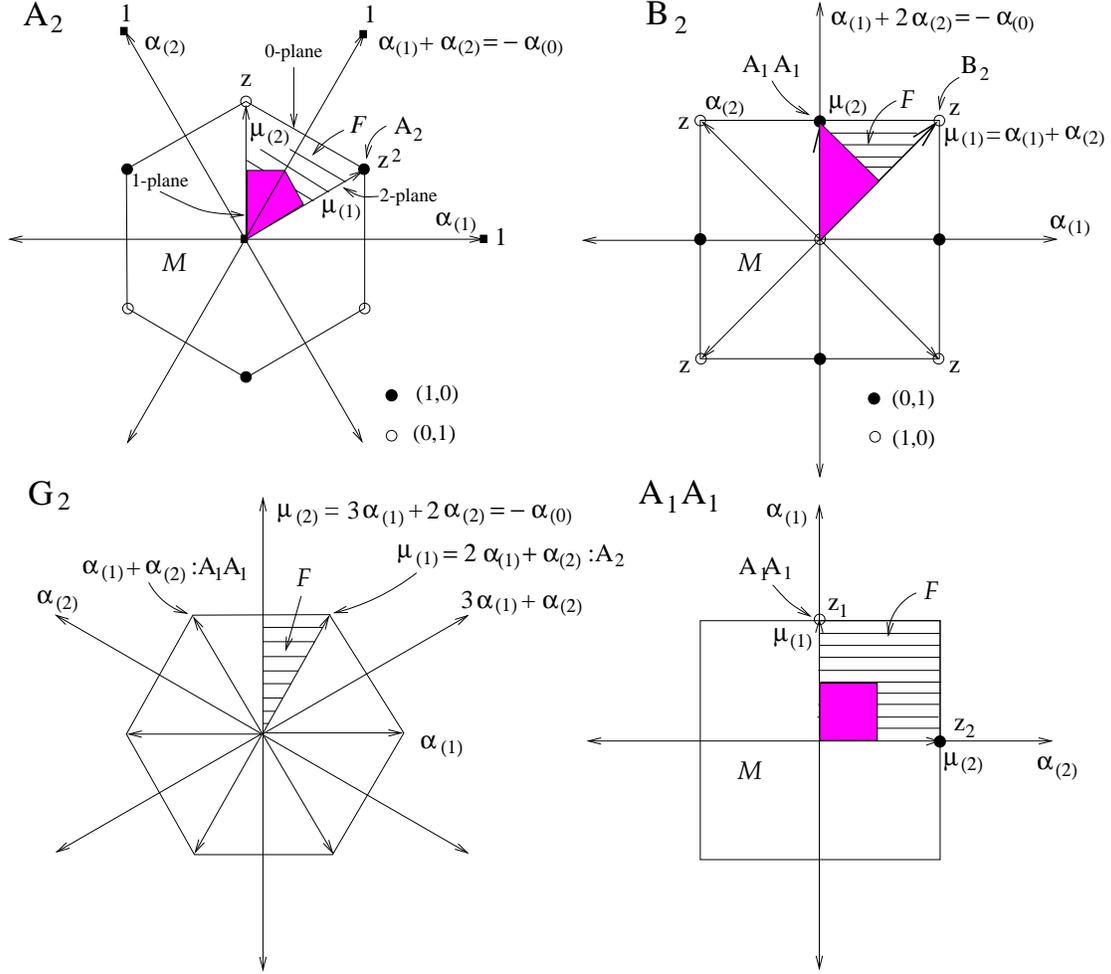}}
\caption{\label{su3}\textsl{
Roots, fundamental weights, center elements,
centralizer subgroups and fundamental domains $\CF$
for the rank 2 case shown. The shaded regions inside
$\CF$ are the fundamental domains for the adjoint
representations.}}
\end{minipage}
\end{figure}

An equivalent definition of $\CF$ is that $\CF$ is the convex set with
extremal points
\eqnl{
\big\{0,{1\ov n_1}\muc_{(1)},{1\ov n_2}\muc_{(2)},\dots,
{1\ov n_r}\muc_{(r)}\big\}.}{wc4}
This can be seen by
expanding $\rho$ 
in terms of the co-weights 
\eqnl{
\CF=\Big\{\rho=\sum_i \xi_i\,\muic\vert\;\;
\xi_i\geq 0,\quad (n,\xi)\leq 1\Big\},}{wc3}
where $n=(n_1,\dots,n_r)$ being the $r$-vector formed
from the Coxeter labels.
For example, the fundamental domains $\CF$ for the $A_r$ and 
$C_r$ groups are the simplicial boxes with extremal points
$\{0,\mui,\;i=1,\dots,r\}$ (recall, that we have chosen
$\vert\al_L\vert^2=2$).
Also, if $\al_1$ and $\al_r$ are the
long and short roots at the endpoints of the $B_r$-Dynkin-diagram,
the fundamental domain for $B_r$ is the convex set 
with extremal points
\eqnn{
\big\{0,\mu_{(1)},\;\ha\mu_{(2)},\;\ha\mu_{(3)},\;\dots 
\ha\mu_{(r-1)},\;\mu_{(r)}\big\}.}
The fundamental domains $\CF$ and the center 
elements for the gauge groups of rank $2$ 
are depicted in fig.\ref{su3}. 
The fundamental domain of $A_2$ is an equilateral triangle,
that of $B_2$ half a square, that of $G_2$ half of an
equilateral triangle and that of $A_1\times A_1$ is
a square. The reflections on the $r$
walls of $\CF$ through $0$ generate the Weyl group $\CW$ of $G$
and give rise to $\CM$.

\begin{figure}[ht]
\begin{minipage}[ht]{16cm}
\centerline{\epsfysize=3.8 cm\epsffile{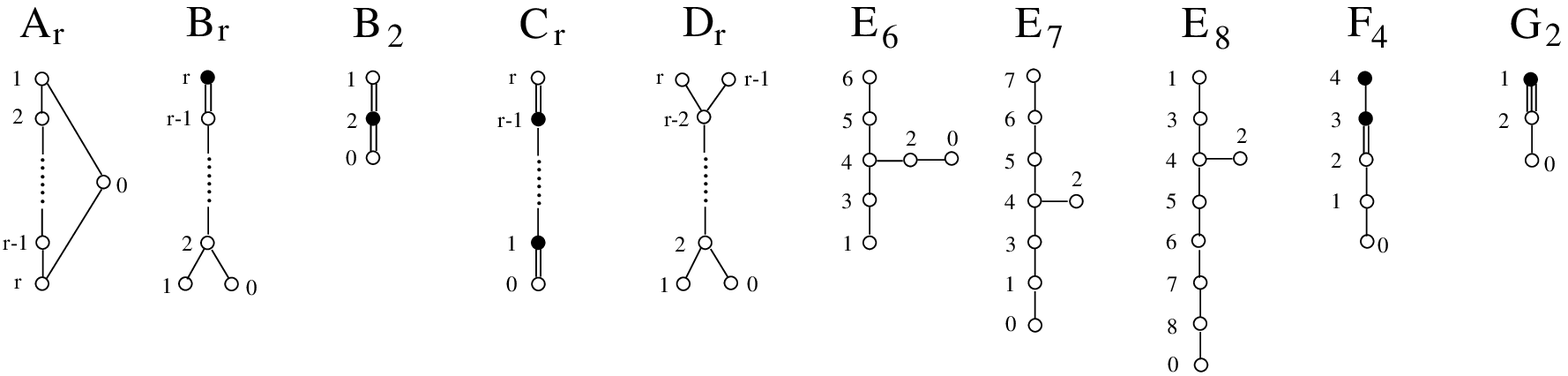}}
\caption{\label{xdynkin}\textsl{
The extended Dynkin diagrams, $\circ$: long roots, $\bullet$: short roots,
$0$: most negative roots
(vertices are labelled as in \cite{bour}).}}
\end{minipage}
\end{figure}

Since $(\alz,\ali)\leq 0$, the highest root $-\alz$ is always
inside the Weyl chamber \refs{weylch} or on its boundary. 
Indeed, for all groups with the exception of $A_2\,$ 
$\,\,-\alz$
lies on  the boundary of $\CF$. From the 
\textit{extended Dynkin diagram}\footnote{One adds the 
most negative root $\alz$ to the
system of simple roots $\ali$ and uses the well-known rules to draw the
Dynkin diagram of this extended system of roots.}(see fig.\ref{xdynkin})
one reads off that for all but the $A_r$ algebras the highest root
is orthogonal to $r-1$ simple roots. Hence it must be proportional
to the weight $\mui$ corresponding to the simple root $\ali$
with $(\ali,\alz)\neq 0$.\pan

Although our strategy is to work  in the covering group
with suitably restricted twists rather than directly
dealing with arbitrary representations, we could in principle do without
twists if we used transition functions and
fundamental domains $\CF_R$ appropriate to the representation $R$.
Actually it is quite straightforward to construct domains
$\CF_R$ for any representation. The volume of such domains
is always less than or equal to that of $\CF$; more precisely
\eqnn{
\hbox{Vol}(\CF_R)={\hbox{Vol}(\CF)\ov \vert\CC_R\vert},}
where $\CC_R$ is the subgroup of the center $\CC$ which is mapped to the 
identity by going from the covering group to the representation $ R $ and 
$\vert\CC_R\vert$ is its order. 
For a given group, the domain with the smallest volume is that for the adjoint
representation since the center is trivial in this case.
The fundamental domains for the adjoint representation
for the rank two groups are shown in figure \ref{su3}.

\mysection{Defects}

Although the Polyakov loop operator itself is smooth for
smooth gauge potentials the factors
$W(\vx)$ and $D(\vx)$ in the decomposition \refs{diagonalization} are in
general not. In this section we shall see that singularities
(so called defects) occur at points $\vx$ at which $h(\vx)$ is 
on the boundary of the fundamental domain $\CF$. At such defects
the residual gauge freedom is enlarged. We shall explicitly determine
the residual gauge groups at the various defects.

From now on we shall assume that $h(\vx)$ is in the
fundamental domain $\CF$. Then 
\refs{diagonalization} assigns a unique $D(\vx)$ (and thus
a unique $h(\vx)\in \CF$) to each Polyakov loop operator since
we have fixed the Weyl reflections.
However, the diagonalizing matrix $W(\vx)$ in 
\refs{diagonalization} is determined only up to right-multiplication with an
arbitrary matrix commuting with  $D(\vx)$
\eqnl{
W(\vx)\longrightarrow W(\vx)V(\vx),\quad 
V(\vx)D(\vx)V^{-1}(\vx)=D(\vx),\qquad
D(\vx)=e^{2\pi i h(\vx)}.}{residual}
At each point the residual gauge transformations $V(\vx)$
form a subgroup of $G$, the centralizer of $D(\vx)$ in 
$G$, denoted by $\CC_{D(\vx)}(G)$. The centralizer contains
the toroidal subgroup of $G$. At points where the
centralizer is just the toroidal subgroup we can
smoothly diagonalize the Polyakov loop operator.\pan
However, at points where the centralizer is non-Abelian
$\CP(\vx)$ has degenerate eigenvalues and
there are obstructions to diagonalizing
$\CP(\vx)$ smoothly \cite{us,tHooft,schierholz}. For what follows it
is useful to define the \textit{defect manifold}
\eqnl{
\CD=\{\vx\in \T^3\vert \CC_{D(\vx)}(G)\neq U^r(1)\}}{defectmani}
on which the centralizer is non-Abelian.
In the special case $G=SU(2)$ the defect manifold is 
$\CD=\{\vx\in \T^3\vert \CP(\vx)=\pm \id\}$.
A \textit{defect} $\CD_i$ is understood to be
a connected subset of $\CD$. In the neighborhood of a
defect the diagonalization is in general not smoothly possible
and the gauge fixing will be singular.
Note that $\CD$ is invariant under time-periodic gauge 
transformations so that the positions of the defects
are gauge invariant.\pan
Now we are going to classify the various defects
which arise in our gauge fixing. To do that we
expand $h(\vx)$ in \refs{residual} into a basis of the Lie algebra
as $h(\vx)=\rho(\vx)\cdot H$ so that
\eqnn{
D(\vx) E_\al D(\vx)^{-1} = e^{2\pi i\big(\rho(\vx),\al\big)}\,E_\al.}
We see that $D(\vx)$ commutes 
with the subgroup $SU(2)$ corresponding to $\al$ if and only if
$(\rho(\vx),\al)$ is integer-valued. 
For $\rho\in\CF$ in \refs{weylch} this can only happen 
if $\rho$ lies on the boundary of the fundamental 
domain. We parametrize $\rho(\vx)$ as in \refs{wc3} so that
\eqnn{
D(\vx) E_{\ali} D(\vx)^{-1} =e^{2\pi i\xi_i(\vx)}\,E_\ali\mtxt{and}
D(\vx) E_{\alz} D(\vx)^{-1} =e^{-2\pi i(\xi(\vx),n)}\,E_{\alz}.}
Therefore $D$ commutes
with the $SU(2)$-subgroup corresponding to the simple roots $\ali$ 
if and only if $\xi_i=0$ and it commutes
with the $SU(2)$-subgroup corresponding to  $\alz$ if and only if 
$(\xi,n)=1$. In other words, the centralizer contains the
$SU(2)$ corresponding to $\als$ if the defect is on
the $\sigma$-plane, i.e. the plane perpendicular to $\als$.\pan
The centralizer of $D(\vx)$ generated by these $SU(2)$ subgroups
can be read off from the extended Dynkin diagram
(see fig.\ref{xdynkin}) as follows: keep the vertex $\sigma$
assigned to the root $\als\in\{\alz,\ali\}$ in the 
extended Dynkin diagram if and only the defect lies on the
$\sigma$-plane. Remove
the other vertices and bonds attached to them. The remaining
diagram is then just the Dynkin diagram belonging to the 
semisimple factor of the centralizer. To obtain the complete
centralizer group
we must multiply with as many $U(1)$-factors as are needed to get
a group of rank $r$.\pan
Let us illustrate how this works for the simply laced groups
$G=A_r$ for which the fundamental domains $\CF$ can
be parametrized as
\eqnn{
\rho=\sum_1^r \xi_i\,\muic,\qquad \xi_i\geq 0,\quad \xi_0\equiv 
1-\sum_1^r \xi_i\geq 0.}
The extremal points of the fundamental domain are
$\{\musc\}$ and they correspond to the 
$r\!+\!1$ center elements of $A_r$. 
If just one $\xi_\sigma$ vanishes then $\rho$ lies inside the 
$(r-1)$-dimensional $\sigma$-plane.
and we must keep the vertex $\sigma$ in the 
extended Dynkin of $A_r$, i.e. the leftmost diagram 
in fig.\ref{xdynkin}. The
corresponding centralizer is  $A_1\times U^{r-1}(1)$. We call such
defects with minimal non-Abelian centralizers 
\textit{basic defects}.
If $\xi_\sigma$ and $\xi_{\sigma^\pr}$ vanish in which
case the defect lies both on the $\sigma$- and $\sigma^\pr$-plane,
then we must keep the two vertices $\sigma$ and $\sigma^\pr$
in the extended Dynkin diagram. If they
are neighbors in figure \ref{xdynkin}, then the centralizer is 
$A_2\times U^{r-2}(1)$,
otherwise it is $A_1\times A_1\times U^{r-2}(1)$.
In the extreme case where just one $\xi_\sigma$ does
not vanish (then $\rho$ is one of the extremal points
of $\CF$) we must retain all vertices with the exception of
the vertex $\sigma$. We get the Dynkin diagram of $A_r$ and
the centralizer is the whole gauge group. By scanning the
whole boundary of $\CF$ comprising of $r\!-\!1$-dimensional,
$r\!-\!2$-dimensional,\dots,$1$-dimensional simplices and the
extremal points we obtain all stabilizer subgroups of $G$.

\mysection{Quantization of the magnetic charges}

In this section we define the Abelian magnetic potential 
$A_{mag}(\vx)$ associated with the partial gauge fixing
and show that the magnetic charge of any defect is
quantized. 
Away from the defects the centralizer of $D(\vx)$ is $U^r(1)$ and
$W(\vx)$ in \refs{diagonalization} is unique up to a residual 
Abelian gauge transformation \refs{residual}:
\eqnl{
W(\vx)\longrightarrow W(\vx)V(\vx)\mtxt{with}
V(\vx)=e^{-i\lam(\vx)}\in U^r(1)\mtxt{on}\CD^c.}{diagonalize}
If we append to each point in $\CD^c$ the set of 
all diagonalizing matrices $W(\vx)$ we obtain a  $U^r(1)$ principal bundle
over $\CD^c$. If we can find a smooth global section in this bundle 
then the diagonalization is smoothly possible outside of the defects, see also
\cite{griesshammer}. 
To investigate the structure of the bundle we employ the
Abelian $U^r(1)$ gauge potential, $A_{mag}(\vx)$,
obtained by projecting the pure gauge $A(\vx)=iW^{-1}(\vx)dW(\vx)$
onto the Cartan subalgebra, i.e.
\eqnn{
A_{mag}(\vx):=A_c(\vx),}
where the subscript $c$ denotes projection onto
the Cartan subalgebra of $G$. 
This potential is singular at the defects and on Dirac strings
joining the defects.
Under a residual gauge transformation \refs{diagonalize}
the gauge potentials transform as
\eqnn{
A_{mag}\longrightarrow A_{mag}+i(V^{-1}dV)_c=A_{mag}+d\lam \mtxt{on}\CD^c.}
Since $A$ is pure gauge the corresponding field strength is given by
\bg\label{mfield}
F_{mag}=\d A_{mag}
=i(A\wedge A)_c,\eg
and it is invariant under residual $U^r(1)$-gauge transformations. 

\pan
Next we will show that a defect may carry $r$ quantized
magnetic charges \cite{thooftpolyakov}. For each defect these charges form
a matrix  $Q_M$ in the Cartan subalgebra $\CH$,
\eqnl{
Q_M= \frac{1}{2 \pi} \int_{\CS} F_{mag} . }{magncharge}
Here $\CS$ is a surface surrounding the defect
$\CD_i$.
\begin{figure}[ht]
\begin{minipage}[ht]{16cm}
\centerline{\epsfysize=5 cm\epsffile{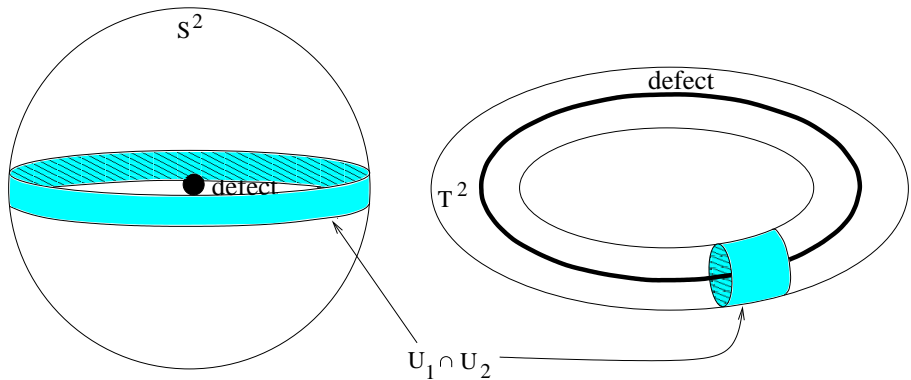}}
\caption{\label{spheretorus}\textsl{
Two typical defects: a monopole and a magnetic loop with surrounding surfaces
and overlap regions.}}
\end{minipage}
\end{figure} 
Excluding walls extending over the whole $3$-torus 
this surface is either a $2$-sphere or a $2$-torus (see
fig.\ref{spheretorus}).
For each $U(1)$ the magnetic charge is just the
instanton number of an Abelian gauge model on $S^2$
or $\T^2$ \cite{waia,sachs} and hence is quantized. More explicitly,
the magnetic charges are the
winding numbers of the map $\exp(i\lam):S^1\longrightarrow U^r(1)$,
\eqnn{
Q_M={1\ov 2\pi}\oint_{S^1}\d\lam,}
where $S^1$ is in the overlap of the two patches
$U_i$ one needs to cover $S^2$ or $\T^2$.
Since the gauge transformation $\exp(-i\lam)$ is
single valued on the overlap,  $Q_M\in\CH$ must satisfy
\eqnl{
e^{2\pi iQ_M}= \id \mtxt{for each defect.}}{quantbed}
For simply connected $G$ this equality must hold on all
states $\vert\mu\rangle$ and we find
\eqnl{
Q_M=\al^\vee\cdot H,\quad\mtxt{where}\al^\vee\in \hbox{co-root
lattice.}}{quantfu} 
Thus we obtain the same magnetic charge quantization
as uncovered by Goddard, Nuyts and Olive \cite{go}
in their pioneering work on electric-magnetic
duality in Yang-Mills-Higgs theories.

\mysection{Instantons and magnetic monopoles}

In this section we work with the simply connected covering
group and exclude twists\footnote{See section 8 where we
included twists for the relevant example $G=SU(3)$.}.
Depending on the residual gauge symmetry in the defects
we get different types of magnetic monopoles.
There are $r+1$ kinds of basic monopoles with minimal
non-Abelian centralizer $SU(2)\times U^r(1)$, corresponding
to the $r+1$ hyperplanes which make up the boundary of the 
fundamental domain.
We will show that a basic defect on the $\sigma$-plane
has magnetic charge
\eqnl{
Q_M=n\,\alsc\cdot H,\qquad\sigma\in\{0,1,\dots,r\}}{magneticcharge}
with integer $n$.
If we have a defect which is on two or more
of the hyperplanes (which means that the Polyakov loop
has more than two degenerate eigenvalues)
then the magnetic charge of this defect
is an integer combination of the co-roots perpendicular
to these hyperplanes.
Below we argue that in general the total magnetic charge
of the defects on a given face gives the instanton number.
For example, the magnetic charge of a defect
on the $0$-plane is
$Q_M=(n\alzc+\beta^\vee)\cdot H,\, n\in \Z$, where
$\beta^\vee$ is in the co-root lattice. This decomposition
of the magnetic charge is unique, see below.
Now the instanton number is simply
\eqnl{
q=-\sum_{\tiny\hbox{defects on 0-plane}} n} {instantonno}
This is our main result. Some illustrative examples
of the use of this formula are given in section 8.
\vskip 0.2truecm\pan
To derive the results (\ref{magneticcharge},\ref{instantonno}) we 
assume that:
\begin{itemize}
\item There are no wall 
defects\footnote{We can formally define
the absence of walls as follows. Consider the extension of the defect
manifold to $\R^3$, i.e.
$\tilde{\cal D}=\{x\in\R^3|{\cal C}_{D(\vx)}\neq U^r(1)\}$
There are no walls if $\tilde{\cal D}^c=\R^3\setminus \tilde D$
is connected.}
\item
Inside a defect the centralizer ${\cal C}_{D(\vx)}$ is uniform.
\end{itemize}
The first assumption is a reflection of the fact that one cannot
surround a wall defect with a closed surface and so it is not
obvious how to define the magnetic charge of such a defect.
The second assumption is made to avoid the complication 
of `defects within defects'. It may be possible to drop this requirement.

Our arguments are based on the observation that
\eqnl{
l\intl_{\T^3}\Tr (\CP^{-1}\d\CP)^3 =
\intl_{\T^3}\Tr \big(P^{-l}dP^l\big)^3}{aaa}
and furthermore
\eqnl{
\Tr \big(P^{-l}dP^l\big)^3 = 
\d\CA^{(\sigma)},\qquad\sigma\in\{0,i\}}{aab} 
where the $2$-forms are
\eqnl{
\CA^{(\sigma)} =-12 l\,\pi i\,\Tr \left[A\wedge A\left(h-{1\ov n_\sigma}\musc
\cdot H\right)\right] + 3\,\Tr\left[A D^{-l}\wedge A D^l \right].}{a}
Here $l$ is the least common multiple of the Coxeter labels $n_i$ and
as before $\mu^\vee_{(0)}\equiv 0$ and $n_0\equiv 1$.
We prove this crucial identity in appendix B.
These 2-forms are  well defined outside the defects, because
they are invariant under the residual Abelian gauge transformations
\refs{diagonalize}.
Both terms in (\ref{a}) may be singular at defects.
However, in the following section we will show that 
$\CA^{(\sigma)}$ can be singular only at defects on the $\sigma$-plane
or equivalently at defects whose centralizers have $\als$ as root,
\eqnl{
\CA^{(\sigma)}\hbox{ singular }\Longleftrightarrow 
\hbox{ defect is on }\sigma\hbox{ plane}\Longleftrightarrow
\als\hbox{ is a root of defect centralizer.}}{regular}

Actually, in \refs{a} we could have subtracted an arbitrary
constant Lie algebra element from $h(\vx)$ and \refs{aaa}
would still hold true. But the smoothness conditions
(\ref{regular}) only hold if
this constant element is an extremal point of the fundamental
domain and if
\eqnn{
\exp{\Big(2\pi i{l\ov n_\sigma}\musc\cdot H\Big)}}
is a center element. Thus we take for $l$ in \refs{aaa}
the least common multiple of the Coxeter labels $n_i$.
For example $l=1$ for the $A_r$ series and $l=2$ for the other
classical groups.\pan

\begin{figure}[ht]
\begin{minipage}[ht]{16cm}
\centerline{\epsfysize=5 cm\epsffile{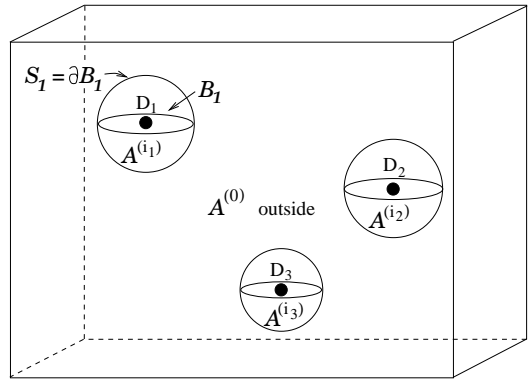}}
\caption{\label{instmon}\textsl{
We must choose two forms $\CA^{(i_p)}$ which are 
regular inside spheres $S_p$ containing a defect on
the inhomogeneous $0$-face.}}
\end{minipage} \end{figure}

Now we make use of \refs{aaa} to relate the magnetic charges
of the defects on the $0$-plane to the instanton number. 
Away from defects on the $0$-plane $\CA^{(0)}$ is regular.
Now we surround each defect $D$ on the $0$-plane with
a closed surface $\CS$ and pick a two form 
$\CA^{(i)}$ which is smooth inside $\CS$, see fig.\ref{instmon}. Since a
defect can lie on at most $r$ of the $r+1$ faces constituting
the boundary of $\CF$ there is always at least one such regular
two form.
With (\ref{topological},\ref{aaa}) the instanton number reads
\eqnl{
\qinst = \frac{1}{24 \pi^2 l} 
\intl_{\scriptscriptstyle \rm{outside}}d\CA^{(0)} +
\frac{1}{24 \pi^2 l} \sum_p\intl_{{\cal B}_p}d \CA^{(i_p)}
=\frac{1}{24 \pi^2 l}\sum_p\intl_{{\cal S}_p}
(\CA^{(i_p)}-\CA^{(0)}),}{instantonnu}
where, since $\CA^{(0)}$ is periodic on $\T^3$,
we get no contributions from the `boundary of the 
torus'\footnote{For twisted gauge fields there are surface
contributions, see section 8.}. Using (\ref{a}) we obtain
\eqnn{
\CA^{(i)}-\CA^{(0)}=
{12\pi i\,l\ov n_i} \Tr \left(A\wedge A \,\muic\cdot H\right).}
Since the magnetic field $F_{mag}$ 
is the projection to the Cartan of $iA\wedge A$ we find
\eqnl{
\CA^{(i)}-\CA^{(0)}={12\pi\,l\ov n_i} \Tr (F_{mag}\, \muic
\cdot H)}{diff}
and end up with
\eqnl{
\qinst= \sum_{D_p}{1\ov n_{i_p}}
\Tr \left(Q_M \,\mu^\vee_{i_p}\cdot H \right),}{qmag}
where we used \refs{magncharge}. The sum extends over
defects on the inhomogeneous $0$-plane.
Let us have a closer look at the contribution 
\eqnl{
{1\ov n_i}\Tr \left(Q_M \,\muic\cdot H \right)}{onedefect}
of a given defect on the $0$-plane.
Consider first a \textit{basic defect} with minimal 
non-Abelian centralizer. Then all two forms $\CA^{(i)},\,i\in\{1,\dots,r\}$ 
are regular and must lead to the same contribution \refs{onedefect}.
We see at once that the magnetic charge must be proportional 
to $\alzc$, 
\eqnn{
Q_M=n\alzc\cdot H,\quad n\in \Z}
and it contributes $n$ to the instanton number.

A non-basic defect on the inhomogeneous face must also lie
on at least one of the homogeneous faces, say the $i$-plane. 
For such a defect we must not take the corresponding singular $\CA^{(i)}$ 
in \refs{instantonnu} or $\muic$ in \refs{onedefect}. 
We see that $Q_M$ may be an integer linear combination of 
$\alzc$ and $\alic$. More generally, if the defect 
lies on the $0$-plane and several homogeneous planes, then
\eqnl{
Q_M=\left(n\alzc+\sum m_i\alic\right)\cdot H,\qquad
m_i\neq 0\mtxt{if defect is not on plane} i.}{chargedec}
Since a defect on the $0$-plane can at most sit
on $r-1$ of the $r$ homogeneous planes, the representation
\refs{chargedec} for the magnetic charge is unique.

Outside of the defects we could have taken any 
$\CA^{(\sigma)}$ instead of $\CA^{(0)}$. Then
only defects on the $\sigma$-plane would contribute to
the instanton number and we would find
\eqnn{
q=\sum_{\tiny\hbox{defects on $\sigma$-plane}}
\Tr \left(Q_M (\mu^\vee_{(\rho)}-\musc)\cdot H \right).}
Again the contribution of a given defect must not depend
on $\rho$ if the corresponding two form $\CA^{(\rho)}$
is regular on the defect. As above we conclude that the 
magnetic charge of a defect is in the co-root lattice
of the defect centralizer,
\eqnl{
Q_M=\left(n\alsc+\sum m_\rho\al_{(\rho)}^\vee\right)
\cdot H,\qquad m_\rho\neq 0\mtxt{if defect is not on plane} \rho,}{chargedec1}
and that the instanton number is 
\eqnn{
q=-\sum_{\tiny\hbox{defects on $\sigma$-plane}} n.}

\def\ll{ l}
\mysection{ Higgs fields}

In this section we consider a parametrization of ${\cal P}(\vx)$ in terms of
static `Higgs' fields.
This may seem to be a backward step since we are encoding a smooth
group-valued object, ${\cal P}(\vx)$, in terms of $r+1$, in general singular,
Lie algebra-valued fields.
However the Higgs fields facilitate a very direct proof
that the ${\CA}^{(\sigma)}$ $2$-forms introduced in the previous section have
the stated smoothness properties.
Moreover, we shall see that the magnetic charges of the defects can be 
related to Higgs winding numbers around the defects.
  
One can define a `basic' Higgs field, $\phi^{(0)}$, as follows
\bg
{\cal P}(\vx)=\exp\left[2\pi i\phi^{(0)}(\vx)\right]\quad\hbox{with}\quad
\phi^{(0)} (\vx)=W(\vx)h(\vx)W^{-1}(\vx).\eg
Now, $\phi^{(0)}(\vx)$, is smooth everywhere except for the inhomogeneous
$0$-plane. This follows because
the centralizer of $D(\vx)$ commutes with $h(\vx)$
unless $(\rho,\alz)=-1$.
It is possible to define `alternative' Higgs fields which
are smooth on the $0$-plane, but singular
on one of the homogeneous $i$-planes, i.e. consider
\bg
\phi^{(i)}=W(\vx)\left(h(\vx)-\frac{1}{n_i}
\mu_{(i)}^\vee\cdot H\right)W^{-1}(\vx),\quad
i=1,2,...,r.\eg
$n_i$ being the $i$'th Coxeter label.
The field $\phi^{(i)}$
is smooth everywhere except points on the $i$-plane.
The relation between the Polyakov loop and the alternative Higgs
fields is as follows
\eqnn{
[\CP(\vx)]^{n_i}z_i=\exp\left[2\pi i n_i \phi^{(i)}(\vx)\right],}
where $z_i$ is the center element $\exp[2\pi i \mu_{(i)}^\vee\cdot H]$.
The $r+1$ Higgs fields $\phi^{(\sigma)},\,\sigma\in\{0,i\}$ 
`cover' the group in the sense that it is possible to partition $\T^3$ into 
patches, so that in each patch at least one of the
Higgs fields is smooth.

In the previous section we wrote $\Tr (\CP^{-\ll}d\CP^\ll)^3$
as the derivative of  two forms $\CA^{(\sigma)}$.
We claimed that $\CA^{(\sigma)}$ is only singular on
the $\sigma$-plane.
In other words, wherever $\phi^{(\sigma)}$ is smooth
$\CA^{(\sigma)}$ is smooth.
This is obvious in the light of the following
identity
\bg\label{mainformula}
\CA^{(\sigma)}=
12\pi^2 \ll^2 \int^1_0
ds (s-1)\Tr\left[
\exp(2\pi is\ll\phi^{(\sigma)})d\phi^{(\sigma)}\wedge
\exp(-2\pi is\ll\phi^{(\sigma)})d\phi^{(\sigma)}\right],\eg
where as before $\ll$ is the least common multiple of the Coxeter 
labels\footnote{One can prove this identity by inserting 
$\phi = W D W^{-1} $ into the integral and comparing with equation \refs{a}.
Alternatively, one can get it from the identity
$\Tr(e^{-\psi}d e^\psi)^3=
3d\left[
\int_0^1 ds\,(s-1)\Tr\left(e^{-s\psi}d\psi\wedge e^{s\psi}d\psi\right)
\right]$.}.

We now show that the magnetic field, $F_{mag}$ can be written in terms
of the Higgs fields.
Using the fields $\phi^{(i)}$ one can construct normalized
Higgs fields $\hat\varphi^{(i)}$ as follows
 \eqnn{
\hat\varphi^{(i)}(\vx)=\phi^{(0)}(\vx)-\phi^{(i)}(\vx)=W (\vx)
\,\frac{\muc_{(i)}}{n_i} \cdot H\, W(\vx)^{-1}.}
In terms of the normalized Higgs fields, the magnetic
fields are
\eqnn{
-\frac{\ll}{n_i}
 \Tr(F_{mag} \mu_{(i)}^\vee\cdot H)= \pi \ll^2 
\int^1_0\!\! 
ds(s\ms 1)\Tr\left[{}\exp\left({2\pi i  s \ll\hat \varphi^{(i)}}\right)
d \hat\varphi^{(i)}\wedge
\exp\left({-2\pi is\ll\hat\varphi^{(i)}}\right)
d 
\hat\varphi^{(i)}\right]{}}
If the Coxeter label $n_i$ is unity, the integral reduces to
\eqnl{
\Tr(F_{mag}\,\mu_{(i)}^\vee\cdot H)= i
\Tr\left( \hat\varphi^{(i)} d\hat\varphi^{(i)}\wedge 
d\hat\varphi^{(i)}\right).}{final}

Let $\CS$ be a closed surface surrounding a defect.
Since the centralizer of $\muc_{(i)}\cdot H$ in $G$ is
$K\times U(1)$, where $K$ is semi-simple, the 
normalized Higgs field
$\hat\varphi^{(i)}$ induces a map from $\CS$ into a coset space
$\CC_i=G/(K\times U(1))$ with $\pi_2(\CC_i)=\Z$.
That is to each normalized Higgs field $\hat \varphi^{(i)}$ there
is \textit{one} associated winding number which can be identified
with $ \Tr\left(Q_M (\CS) \muc_{(i)}\cdot H\right)$.

For $SU(N)$ all the Coxeter labels are unity, and so
\eqnn{
F_{mag}= i \sum_{i=1}^{N-1}\alpha_{(i)}\cdot H\,\Tr
\left( \hat\varphi^{(i)} d\hat\varphi^{(i)}\wedge 
d\hat\varphi^{(i)}\right).}
For the groups $B_r$, $C_r$, $D_r$, $E_6$ and $E_7$
it seems 
that the magnetic field cannot be written trilinearly
in normalised fields since (\ref{final}) only applies 
if the relevant Coxeter label is one.
For example the gauge group $E_7$ has only one unit Coxeter label, $n_7$.
However, the Weyl orbit of $\mu_{(7)}^\vee$ contains a linearly
independent basis of  the root space.
To make this more concrete, consider the  field
\eqnn{
\hat\varphi_X= W(\vx)\,X\cdot H\, W^{-1}(\vx).}
A simple calculation shows that
\eqnn{
\Tr(F_{mag} X\cdot H)={i}
\Tr\left( \hat\varphi_X \,d\hat\varphi_X\wedge 
d\hat\varphi_X\right),}
if and only if
\eqnl{
(X,\alpha)^3=(X,\alpha) \quad\mtxt { for all roots $\alpha$.}}{cubic}
Clearly, $X=\mu_{(i)}^\vee$ is
a solution of (\ref{cubic}) if and only if
$n_i=1$. But there are other solutions of (\ref{cubic}) apart from
the co-weights with unit Coxeter; these correspond to Weyl reflections of
the co-weights. In fact for $B_r$, $C_r$, $D_r$,
$E_6$ and $E_7$ one can always find $r$ linearly independent solutions
of (\ref{cubic}) which we denote by $X_i,\;i=1,2,...,r$.
Thus we have
\eqnn{
F_{mag}=i\sum_{i=1}^r Y^i\cdot H\;  
\Tr\left(\hat\varphi^{(i)}\, d\hat\varphi^{(i)}\wedge 
d\hat\varphi^{(i)}\right),}
where now $\hat\varphi^{(i)}=\varphi_{X_i}$, and the 
$Y^i$ are dual to the $X_i$ in the  sense that
$(Y^i,X_j)=\delta^i_{\,\,j}$ (the $Y^i$ are roots or Weyl reflections
thereof). 
To each normalized Higgs field $\hat \varphi^{(i)}$ there
is \textit{one} associated winding number which can be identified
with $ \Tr\left(Q_M\,X_i\cdot H\right)$.

For the groups $E_8$, $F_4$ and $G_2$ no solutions of
(\ref{cubic}) exist.

\mysection{$SU(3)$} 
\label{examples}

In this section we illustrate  the ideas of the previous
sections by considering the relevant gauge group $SU(3)$.
In the instanton number calculation of chapter 6
we assumed that our matter transformed according to the covering group.
Here we will also consider the case of matter in the adjoint
representation by allowing for twists.
 
\begin{figure}[ht]
\begin{minipage}[ht]{16cm}
\centerline{\epsfysize=4.5 cm\epsffile{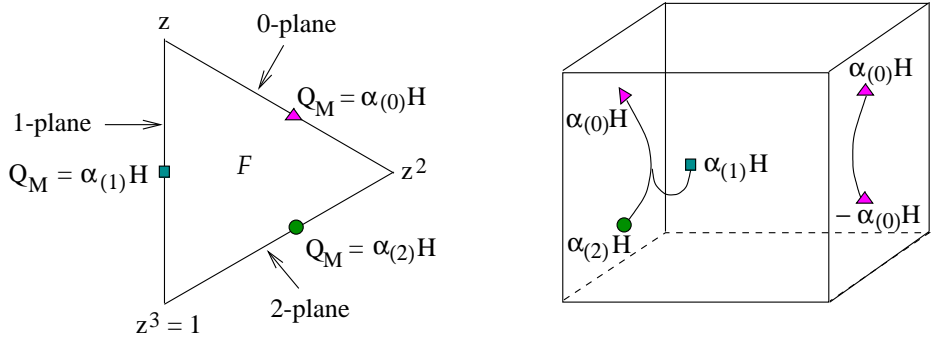}}
\caption{\label{bild5}\textsl{
The fundamental domain $\CF$ for $G=A_2$ with elementary magnetic charges
corresponding to the defects and a string network connecting
different basic monopoles. Shown is a network with instanton number $-1$}}
\end{minipage} \end{figure}

First we consider $SU(3)$ with untwisted gauge fields, i.e.
the Polyakov loop operator in the defining representation.
The fundamental domain $\CF$ has been
depicted in figs.(\ref{su3}a,\ref{bild5}). The magnetic charges of the
three types of defects corresponding to the three edges
of $\CF$ are integer multiples of
\eqnn{
\al^\vee_{(1)}\cdot H=\pmatrix{1&0&0\cr 0&-1&0\cr 0&0&0},\quad
\al^\vee_{(2)}\cdot H=\pmatrix{0&0&0\cr 0&1&0\cr 0&0&-1},\quad
\alzc\cdot H=\pmatrix{-1&0&0\cr 0&0&0\cr 0&0&1}}
Because of overall charge neutrality the magnetic charges of all
defects must add up to zero,
\eqnn{
\sum_{\rm{all\ defects}}\;Q_M=0.}
Any cluster of magnetic monopoles connected by a Dirac string
has vanishing magnetic charge. For example, if a monopole pair is uncharged
no Dirac string, besides the one connecting the two monopoles,
is needed. Since defects on the $0$-plane for which
$Q_M=n\,\alzc\cdot H$ (ignoring `higher defects') 
contribute to the instanton number as
\eqnn{
\qinst=\sum_{\tiny\hbox{defects on $0$-plane}}
\Tr\Big(Q_M\,\mu_{(1)}^\vee \cdot H\Big)}
the monopole pair connected by a string in fig.\ref{bild5}
does not contribute to the instanton number. The three monopoles connected by
a Dirac string contribute $-1$ to the instanton number.

What about defects with larger centralizers?
If $\CP(\vx)=z$, in which case $h(\vx)$ lies at
an extremal point of $\CF$ in fig.\ref{bild5},
then $\CP$ has maximal degeneracy and the centralizer is $A_2$.
Such a defect has magnetic charge
\eqnn{
Q_M=n_1\,\al^\vee_{(1)}\cdot H+n_2\,\alzc\cdot H=
\Big((n_1-n_2)\al^\vee_{(1)}-n_2\al^\vee_{(2)}\Big)\cdot H,\qquad
n_i\mtxt{integers}}
and contributes with $-n_2$ to the instanton number.

Finally, let us switch to the adjoint representation.
In principle we could do this by restricting $h(\vx)$ to the fundamental
domain for the adjoint representation, see the 
shaded regions in fig.\ref{su3}.
However this would lead to walls on which $ W(x) $ is not smooth.
A much easier approach is to  work in the covering group
\textit{ but now with arbitrary twists.}
In general this leads to a fractional instanton number.
Such fractional instanton numbers are related to a loss
of charge neutrality and nonperiodicity of $ \CP(\vx) $ 
engendered by the twists.

\begin{figure}[ht]
\begin{minipage}[ht]{16cm}
\centerline{\epsfysize=3.5 cm\epsffile{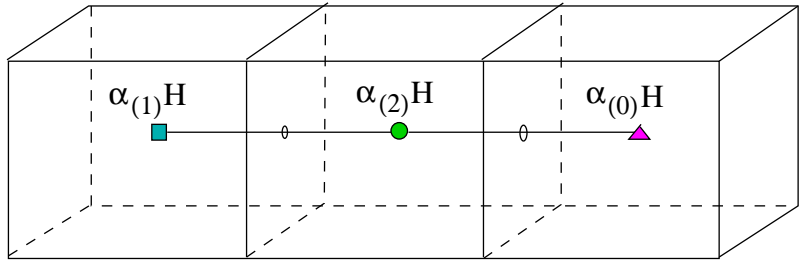}}
\caption{\label{b2}\textsl{
In the twisted sector with $q=1/3$ there maybe just one
basic monopole in the torus. In the tripled torus we have
charge neutrality.}}
\end{minipage}
\end{figure}

For example consider the following set of twists
$z_{01}=\exp[4\pi i/3] \id =\exp[2 \pi i \mu_{(1)}^\vee \cdot H],
z_{23}=\exp[2\pi i/3] \id = \exp[2 \pi i \mu_{(2)}^\vee \cdot H] $, 
and all other twists the identity. This is an example of non-orthogonal twists, 
and leads to an instanton number of the form $q=\frac{1}{3}+n$ where $n\in \Z$.
From the periodicity properties of $ \CP ( \vx ) $ 
\eqnn{
\CP(\vx+b_i) = z_{0i} \omega_i^{-1} \CP(\vx) \omega_i \, , \quad i=1,2,3 }
we obtain periodicity properties of $ W(\vx) $, $ D(\vx)=\exp[2 \pi i h(\vx)] $ 
and $ h(\vx) $. In our example we get
\eqnn{
h(\vx+b_1) = \w \left( h(\vx) - \mu_{(2)}^\vee \cdot H \right) \w^{-1} ,}
where $ \exp ( - 2 \pi i \mu_{(2)}^\vee \cdot H ) = z_{01} $ and $ \w $
corresponds to an element of the Weyl group, here a rotation of $ 2 \pi /3$. 
The equation can be understood as
follows. By multiplying $ D(\vx) $ with $ z_{01} $ we shift $ h ( \vx ) $ 
by the
vector $ - \mu_{(2)}^\vee \cdot H $. Then we have to Weyl reflect this shifted
vector back into the fundamental domain $ \CF $ with $ \w $. In $ \CF $ itself
this corresponds to a rotation with angle $ 2 \pi /3 $ around the center of
the equilateral triangle $ \CF $. It follows that we get charge neutrality 
in the `tripled' torus obtained by taking three adjoining tori in the 
$x_1$-direction. If we have in the first torus a defect of one type 
then in the adjoining torus in the $ x_1 $ direction
we have a defect with the next 
type of charge and so on, see fig.\ref{b2}.
In the $ x_2 $- and $ x_3 $-directions $ h(\vx) $ is periodic 
($z_{02}=z_{03}=\id$). The periodicity
properties of $ W ( \vx ) $ are given by
\eqnn{
W (\vx+b_1) = W ( \vx ) \w^{-1} R_1 (\vx) \quad \mbox{and} \quad
W(\vx + b_i) = \omega_i W(\vx) R_i(\vx),\quad i=2,3 ,}
where $\omega_1$ and $\omega_2$
are twist eaters such that $ \omega_2 \omega_3 = 
\omega_3 \omega_2 z_{23} $ and $ R_i $ are functions with values in the Cartan
subgroup\footnote{In general the functions $ R_i $ can not be chosen smooth on
the whole torus.}. From these conditions we obtain periodicity of 
the magnetic field strength $ F_{mag} = i A \wedge A $ in the $x_2$- and 
$x_3$-directions and $ F_{mag} ( \vx + b_1 ) = \w F_{mag} ( \vx ) \w^{-1} $. 
To calculate the topological index $ q $ we may use the 
2-forms $ \CA^{(\sigma)} $, but now we will get 
contributions from the `boundary'
of the torus. This is in contrast to the non twisted case 
where we have had no contributions from the boundary 
because of the periodicity of $ \CP(\vx)$. 
We assume that there are no defects on the boundary. Then we can integrate 
$ \CA^{(0)} $ over the boundary. One easily checks that $ \CA^{(0)} $ is
periodic in the $x_2$- and $x_3$-directions. Therefore we end up with
\begin{eqnarray}
\nonumber
q_b = \frac{1}{24 \pi^2} \int_{\partial \T^3}  \CA^{(0)} &=& 
\frac{1}{24 \pi^2} \int_{x_1=0} \CA^{(0)} (\vx + b_1) - \CA^{(0)} (\vx) =
\frac{1}{2 \pi} \int_{x_1=0} \Tr(F_{mag} \mu_{(2)}^\vee \cdot H ) .
\end{eqnarray}
This  shows the relation between the noninteger boundary
contribution\footnote{By writing $ F_{mag} = d A $ and using the 
cocycle condition for $ R_2 $ and $ R_3 $ one easily sees that $ q_b $ 
is indeed noninteger.} to the instanton number and the total 
magnetic flux through 
the torus which results from the loss of charge neutrality on
$\T^3$. In our example the element $ \w $ of the Weyl group is a rotation of 
$ 2 \pi / 3 $ in the Cartan subalgebra. Therefore $ r^3 = \id $ which shows 
together with the periodicity properties of $ F_{mag} $ 
that in the tripled torus we have no boundary contributions to the topological
index.

\mysection{Conclusions}

In this paper we have considered gauge-fixing of Yang-Mills theory 
on the four torus for arbitrary gauge groups,
instanton sectors and twists.
We have generalized our earlier results \cite{us,mpw}
on the extended Abelian projection with respect to
the Polyakov loop operator on the four torus.
We have constructed a complete set
of non-Abelian transition functions, which encode
the `boundary conditions' for the gauge potentials, for all instanton
sectors and arbitrary twists. 
With these transition functions the path ordered exponential,
$\CP(\vx )$, which is central to the gauge fixing, is  periodic
up to multiplication by constant matrices, 
even though of course the gauge field itself is non-periodic.
Then we found an explicit
gauge transformation which transforms $A_0$ into the Cartan subalgebra
and hence the Polyakov loop operator into the toroidal subgroup of $G$.
The resulting gauge fixed $A_0$ is time independent. We have
fixed the freedom in choosing the gauge transformation 
by restricting $A_0$ to a fundamental domain in the Cartan subalgebra. 

In the sectors with non-vanishing instanton number the final gauge
fixed potential  must have singularities \cite{tHooft}.
These singularities are due to ambiguities in the diagonalization
of $\CP(\vx )$ at points where the centralizer of $\CP(\vx )$ 
is non-Abelian. There is a close analogy between these defects and magnetic
charges in Yang-Mills-Higgs theories. The defects are classified
according to the non-Abelian centralizer subgroups of $\CP(\vx)$. 
A point $\vx$ belongs to a defect if the gauge fixed $A_0(\vx)$
lies on the boundary of the fundamental domain.  
Here the results for $SU(2)$ may be misleading; at the defects the
Polyakov loop operator need not be in the center of the gauge group
as it must for $SU(2)$. For example, 
for $G\in\{E_8,F_4,G_2\}$ the center is trivial but there are 
many different types of defects corresponding to the
different faces of the fundamental domain. 
The magnetic charges
of the defects are quantized and  linearly related to the points of
the integral co-root
lattice.   For all groups with nontrivial
centers we have constructed $r$ normalized Higgs fields which wind around
the magnetized defects. 
Finally we generalized earlier results in \cite{Jackiw,pisarski,reinhardt,us}
and related the magnetic charges of a given type of defect
to the instanton number $\qinst$. In particular, if $\qinst\neq 0$
then all possible magnetic defects must appear.

One may view our gauge fixing as the `nearest' fixing
to the Weyl gauge compatible with time periodicity.
Yet unlike the Weyl gauge we find monopole like singularities.
This is gratifying, since
in those theories where we analytically understand confinement,
the latter is due to the condensation of monopoles; these
examples are compact $QED$ \cite{poly} and supersymmetric
Yang-Mills theories \cite{seiberg}.
Of course, there
is a long way from the picture of condensed magnetic
monopoles to real $QCD$.

The treatment given here has been purely classical. The next
step would be to study the path integral within this gauge fixing.
At this point one would need a sensible approximation
\cite{lenz}. The balancing of the
energy and the entropy of monopoles (and/or loops) may
explain the occurrence of the deconfinement transition
in $QCD$. 
It would be interesting to clarify the role of the center of the
gauge groups. There are gauge groups with trivial centers
but many different types of monopoles and other magnetic
defects.

\section*{Acknowledgements}

We are grateful to Falk Bruckmann, Jan Pawlowski and Hugo Reinhardt
for helpful discussions. We thank R. Jackiw for bringing \cite{Jackiw}, 
in which the relation between magnetic charges and the instanton 
number has been discovered, to our attention.

\appendix

\section{Transition functions and twist}

We prove that for arbitrary twists and instanton number
there exist transition functions with the following property
\bg
U_0=\id,\quad U_i(x^0=0)=\omega_i,\eg
where the $\omega_i$ are twist eaters satisfying
\bg\label{hcocycle}
\omega_i \omega_j=z_{ij}\omega_j \omega_i,\quad z_{ij} \in {\cal{Z}} .\eg
We now start off with \sl Abelian  \rm
transition functions
\eqnl{
U_\mu=\exp\left[2\pi i\sum_{\nu=0}^3
\frac{{\bf n}_{\mu \nu}x^\nu}{L_\nu}\right],}{Abelian}
where 
${\bf n}_{\mu\nu}$ is a Cartan sub-algebra valued lower triangular
matrix
\bg
{\bf n}_{\mu\nu}=\btensor{(}{cccc}
0&0&0&0\\
{\bf n}^1&0&0&0\\
{\bf n}^2&{\bf m}^3&0&0\\
{\bf n}^3& -{\bf m}^2  &{\bf m}^1&0\etensor{)}.\eg
With this choice of ${\bf n}_{\mu\nu}$ we have $U_0=\id$.
The cocycle condition ensures that the ${\bf n}^i$ and ${\bf m}^i$
satisfy the constraints
\bg
e^{2\pi i {\bf n}^i}=z_{0i},\quad
e^{2\pi i {\bf m}^1}=z_{23}\quad\hbox{and cyclic permutations}.\eg
The instanton number is simply
\eqnn{
\qinst=  
\Tr\,({\bf n}^1 {\bf m}^1+{\bf n}^2 {\bf m}^2+{\bf n}^3 {\bf m}^3).}
Now we claim that there exists a time-independent gauge transformation
$V(\vx )$ with the following properties
\eqnl{
V^{-1}(\vx)U_i(x^0=0,\vx)V(\vx+b_i)=\om_i.}{vprop}
To prove this consider the following two sets of transition functions.
Firstly take the Abelian transition functions
(\ref{Abelian}) but with the ${\bf n}^i$ all set to zero. Secondly
take the set of transition functions
$U_0=\id,\quad U_i=\omega_i$, where the $\omega_i$ are defined as in
(\ref{hcocycle}).
Now both sets of transition functions
have instanton number zero and an identical set of
(magnetic) twists. Hence they must be gauge equivalent \cite{vanbaal}.
This establishes the existence of a smooth $V(\vx )$
satisfying (\ref{vprop}).
Now  we perform this gauge transformation
on the original Abelian transition functions
(i.e. with the ${\bf n}^i$ not necessarily zero)
\bg
U_0^V=\id,\quad
U_i^V=V^{-1}(\vx )\exp\left[2\pi i\sum_{\nu=0}^3
\frac{ {\bf n}_{i\nu}x_\nu}{L_\nu}\right]V(\vx +b_i).\eg
These transition functions have the stated properties.
\pan
This proof hinges on two assumptions:
\begin{itemize}
\item The existence of Abelian transition functions
for arbitrary twists and instanton number.
\item The existence of twist eaters for
all possible magnetic twists $z_{ij}$.
\end{itemize}
It is well known that the first assumption breaks down
in the odd instanton sectors of untwisted
$SU(2)$ gauge theory. This special case has been addressed
in ref. \cite{us}.
In \cite{ambjorn} it was shown that the second assumption
is valid for $SU(N)$.
We will show the existence of magnetic twist eaters also for the other 
simple Lie groups. 
For every group (with the exception of the $D_{2r}$-series, which
will be considered separately) the cyclic center is generated by
\eqnn{
z=\exp\Big(2\pi i\muc_{(z)}\cdot H\Big).}
In the table below we list the co-weights $\muc_{(z)}$ 
generating the centers.
We now argue that magnetic twist eaters can be constructed from
an Abelian element $ A $ 
and an element $ \w $ in the Weyl group.
The Abelian element $ A $ is given by 
\eqnl{
A=\exp\left[\frac{2\pi i}{g} \delta_w\cdot H\right],}{AbelianA}
where $g=1+\sum n_i$ is the Coxeter number (see the table below)
and $\delta_w$ is the Weyl vector
\eqnn{
\delta_w=\sum_i\mu_{(i)}=\ha\sum_{\al>0}\al \, ,\qquad
|\delta_w|^2={\hbox{dim}G\ov 24}g\,|\al_L|^2.}
The element $\w$ is fixed by the requirement that 
\eqnl{
\w^{-1}
(\delta_w\cdot H) \w =\delta_w\cdot H-g\mu_{(z)}\cdot H.}{weylrefll}
Such a Weyl group element $\w$ exists for \textit{all} groups.
For example for $G=SU(N)$ and $\mu_{(z)}=\mu_{(r)}$ it is
\eqnn{
\w=\w_1\w_2\dots \w_{N-1},}
where $\w_i$ is the fundamental reflection on the plane orthogonal
to the simple root $\ali$,
\eqnn{
\w_i^{-1}(\mu\cdot H) \w_i=\sigma_{\ali}(\mu)\cdot H.}
The Weyl word $\w_1\w_2$ first reflects on the plane orthogonal
to $\al_{(1)}$ and then on the plane orthogonal to $\al_{(2)}$.
$A$ and $\w$ have the basic property
\eqnn{
\w^{-1}A \w =z^{-1}\,A \mtxt{so that}\w^{-p}A^q\w^p=z^{-pq}A^q.}
To prove this property we first note, that we may
replace the weight $\mu_{(z)}$ in \refs{weylrefll} by
the corresponding co-weight, since $\al_{(z)}$
is always a long root. Now we conclude that
\eqnn{
\w^{-1}A\w=\exp\left[
{2\pi i\ov g}\w^{-1}\delta_w\cdot H \w\right]=
\exp\big(-2\pi i\muc_{(z)}\cdot H\big)\,A=z^{-1}A,}
as required. 

Now we prove that for given magnetic twists $ z_{ij} = 
z^{\eps_{ijk}t_k},\;t_k\in\Z$ we can find 
twist eaters $ \omega_i$ satisfying equation \refs{hcocycle}.
We make the ansatz
\eqnn{
\omega_i = \w^{p_i} A^{q_i}\mtxt{such that}
\omega_i \omega_j = z^{p_i q_j - p_j q_i}\om_j\om_i .}
It follows that equation \refs{twisteaters} is equivalent to
\def\vn{\vec{n}}
\def\vp{\vec{p}}
\def\vq{\vec{q}}
\eqnl{
\vn\equiv  \vp\wedge\vq\;\;\;\mbox{mod}(\vert\CZ\vert),}{cong0}
where $\vert\CZ\vert$ is the order of the center group.
If all twists are the identity (all $n_i$ are zero) 
the solution is trivial. So let us assume that at
least one $n_i$, say $n_3$ is not zero. Then we choose
\eqnn{
\vp=\pmatrix{0\cr 1\cr p},\quad \vq=\pmatrix{-n_3\cr 0\cr n_1}
\mtxt{so that}
\vn=\pmatrix{n_1\cr -pn_3\cr n_3}.}
It remains to be shown that for a given $n_2$ and $n_3\neq 0$
we can solve
\eqnl{n_2=-pn_3\;\hbox{mod}(\vert\CZ\vert).}{eee}
If the order of the center is a prime number,
as it is for all but the $A$ and $D$ groups, then we can
always find a $p$ solving this equation.
For the $D_r$ groups with odd $r$ the order of the
center is not prime but $4$. If only one $n_i$, say again $n_3$ is odd then 
we can again solve \refs{eee}. In the other case all $n_i$ 
must be even and \refs{eee} can again be solved. This proves 
the existence of twist eaters for all but the $D_{r}$-groups with
even rank.

For the $D_{r}$-groups with even rank the center comprises of
\eqnn{
\id,\quad z_1=e^{2\pi i \mu_{(1)}^\vee\cdot H},\quad
z_2=e^{2\pi i \mu_{(r)}^\vee\cdot H}\mtxt{and}
z_3=e^{2\pi i\mu_{(r-1)}^\vee\cdot H},}
where $z_iz_j=\delta_{ij}\id+\eps_{ijk}z_k$. 
As before one can find commuting Weyl words $\w_{(i)}$
such that for each center element
\eqnl{
\w_{(i)}^{-1} A \w_{(i)} =z_i^{-1} A=z_iA \mtxt{and} 
\w_{(i)} \w_{(j)} =  \w_{(j)} \w_{(i)} .}{twistss}
For example,
\eqnn{
\w_{(1)}=\w_1\w_2\cdots \w_{2r}\w_{2r-2}\w_{2r-3}\cdots \w_1.}
Now we make a case by case analysis to show the existence 
of twist eaters for arbitrary given twists. Using \refs{twistss}
one finds the following solution for the possible choices for
$z_{ij}$ in \refs{hcocycle}:
\begin{center}
\begin{tabular}{|l||c|c|c||c|c|r|}\hline
case 
& $\om_1$& $\om_2$&$\om_3$
&$z_{12}$&  $z_{13}$& $z_{23}$
\\ \hline\hline
one twist&
$A$&$\w_{(i)}$&$\id$&$z_i$& $\id$&$\id$\\ \hline
two twists&
$A$&$\w_{(i)}$&$\w_{(j)}$ 
&$z_i$&$z_j$&$\id$
\\ \hline
$3$ different twists&
$\w_{(i)} A$&$\w_{(j)}A$&$\w_{(k)}A$
&$\eps_{ijk}z_k$&$\eps_{ikj}z_j$&$\eps_{jki}z_i$
\\ \hline
$2$ or $3$ identical twists&
$\w_{(i)}$&$\w_{(j)}A$&$A$
&$z_i$&$ z_i$&$z_j$
\\ \hline
\end{tabular}
\end{center}
Together with the result in \cite{ambjorn} this finally
proves the existence of magnetic twist-eaters for all gauge groups.\pan
In the main body of the paper we needed the 
centers, (dual) Coxeter labels and Coxeter numbers
of the various gauge groups. For completeness we have listed
these in the tables below.\pan
\begin{center}
\begin{tabular}{|l||c|c|c|c|c|}\hline
group& $A_r$  &$B_r$   &$C_r$  &$D_r,\;r\hbox{ even}$  & $D_r,\;r\hbox{ odd} 
$  \\ \hline
${\cal Z}$ & $\Z_{r+1}$ & $\Z_2$ & $\Z_2$ & $\Z_2\times \Z_2$ & $\Z_4 $
\\ \hline
$\muc_{(z)}$&$\muc_{(1)}$&
$\muc_{(1)}$&$\muc_{(r)}$&$\muc_{(1)},\muc_{(r)}$ &$\muc_{(r)}$\\ \hline
$n_i$&$1,\dots ,1$&$1,2,\dots ,2,2$&$2,\dots ,2,1$& 
$1,2,\dots ,2,1,1$&$1,2,\dots ,2,1,1$\\ \hline
$n^\vee_i$& &$1,2,\dots ,2,1$&$1,\dots ,1,1$&
& \\ \hline
$g$&$r+1$ & $2r$ &$2r$&$2r-2$&$2r-2$\\ \hline
\end{tabular}
\end{center}
Table 1a: \textit{
Centers ${\cal Z}$, generators $\muc_{(z)}$ of the centers:
$z=\exp(2\pi i\muc_{(z)})$,
Coxeter labels $n_i$, dual Coxeter labels $n^\vee_i$ and  
Coxeter number $g$ of the classical groups}
\vskip 0.5truecm

\begin{center}
\begin{tabular}{|l||c|c|c|c|c|}\hline
group&  $E_6$& $E_7$& $E_8$& $F_4$&$G_2$\\ \hline
${\cal Z}$& $\Z_3$&$\Z_2$&$\id$&$\id$&$\id$\\ \hline
$\muc_{(z)}$&$\muc_{(1)}$&$\muc_{(7)}$& & & \\ \hline
$n_i$&$1,2,2,3,2,1$&$2,2,3,4,3,2,1$&
$2,3,4,6,5,4,3,2$&$2,3,4,2$&$3,2$\\ \hline
$n^\vee_i$&&&&$2,3,2,1$&$1,2$\\ \hline
$g$& $12$ &$18$&$30$&$12$&$6$\\ \hline
\end{tabular}
\end{center}
Table 1b: \textit{Centers ${\cal Z}$, generators $\muc_{(z)}$ of the centers:
$z=\exp(2\pi i\muc_{(z)})$,
Coxeter labels $n_i$, dual Coxeter labels $n^\vee_i$ and 
Coxeter number $g$ of the exceptional groups}
\vskip 0.5truecm

\section{Proof of \refs{aaa}}

For $\ll=1$ this formula is easily checked if one
uses $\CP=WDW^{-1}$ and the definitions $A=iW^{-1}dW$
and $\log D=2\pi ih$. To prove the formula for $\ll>1$
is less trivial. As a first step consider two group
valued fields $P_1,\,P_2$. Then 
\eqnn{
\Tr\Big((P_1P_2)^{-1}d(P_1P_2)\Big)^3=\sum_i\Tr(P_i^{-1}dP_i)^3
-3\,d\,\Tr\big(P_1^{-1}dP_1\wedge dP_2P_2^{-1}\big).}
If the $P_i$ are smooth and periodic then
\eqnn{
\intl_{\T^3}Tr\Big((P_1P_2)^{-1}d(P_1P_2)\Big)^3=
\sum_i\intl_{\T^3}\Tr(P_i^{-1}dP_i)^3.}
With our choice for the transition functions the Polyakov
loop operator is indeed periodic and we conclude
that
\eqnl{
\intl_{\T^3} \Tr\Big(\CP^{-l}d(\CP^l)\Big)^3=
l\intl_{\T^3} \Tr(\CP^{-1}d\CP)^3.}{powers}
Now we can relate the instanton number in \refs{topological}
to the winding of $\CP^l$ as follows
\eqnn{
q={1\ov 24\pi^2 l}\intl_{\T^3} \Tr \Big(\CP^{-l}\,d\CP^l\Big)^3}
Since $\CP^\ll=WD^\ll W^{-1}$ we can now apply
formula \refs{aaa} with $D$ replaced by $D^\ll$. This
then leads to
\eqnn{
q=\sum_\sigma {1\ov 24\pi^2 l}\intl_{M_\sigma} d\CA^{(\sigma)}\, , \quad 
\bigcup_\sigma M_\sigma = \T^3 \, , \quad 
M_\sigma \cap M_{\sigma '} = \emptyset\, , \, \mbox{if} \, 
\sigma \neq \sigma ', }
where $\CA^{(\sigma)}$ is smooth in $ M_\sigma $ and has been defined in 
\refs{a}. 
This proves \refs{aaa} for $\ll>1$ as required.


\begin{thebibliography}{99}
\bibitem{tH} G. 't Hooft, Nucl. Phys. B153 (1979) 141;
Acta Phys. Austria CA Suppl. XXII (1980) 1063;
Phys. Scr. 24 (1981) 841


\bibitem{pisarski} D.J. Gross, R.D. Pisarski and L.G. Yaffe, Rev. Mod.
Phys. 53 (1981) 43

\bibitem{kapusta}
J.I. Kapusta, \textit{Finite-temperature field theory},
Cambridge University Press, 1989


\bibitem{luesch} M. L\"uscher, Phys. Lett. 118B (1982) 391

\bibitem{wilson} K.G. Wilson, Phys. Rev. D10 (1974) 2445

\bibitem{polloops} A.M. Polyakov, Phys. Lett. 72B (1978) 477;
L. Susskind, Phys. Rev. D20 (1979) 2610

\bibitem{us} C. Ford, U.G. Mitreuter, T. Tok, A. Wipf
and J.M. Pawlowski, 
\textit{Monopoles, Polyakov loops and gauge fixing on the torus},
preprint FSUJ-TPI-98-03, hep-th/9802191, Annals Phys., in press

\bibitem{mpw} U. G. Mitreuter, J. M. Pawlowski
and A. Wipf, Nucl. Phys. B514 (1998) 381

\bibitem{tHooft}G. 't Hooft, Nucl. Phys. B190 (1981) 455


\bibitem{schierholz} A.S. Kronfeld, G. Schierholz and
U.J. Wiese, Nucl. Phys. B293 (1987) 461

\bibitem{dirac} P.A.M. Dirac, Proc. Roy. Soc. A133 (1931) 60

\bibitem{Jackiw} N. Christ and R. Jackiw, Phys. Lett. 91B (1980) 228

\bibitem{reinhardt} H. Reinhardt, Nucl. Phys. B503 (1997) 505

\bibitem{Lenz3} O. Jahn and F. Lenz, Phys. Rev. D58 (1998) 085006

\bibitem{arroyo} A. Gonzalez-Arroyo, \textit{Yang-Mills Fields on
the 4-dimensional torus. Part I: Classical Theory}, preprint FTUAM-97/18,
hep-th/9807108

\bibitem{ambjorn} J. Ambjo$\!\!\!/$rn and
H. Flyvbjerg, Phys. Lett. 97B (1980) 241

\bibitem{vanbaal}P. van Baal, Comm. Math. Phys. 85 (1982) 529

\bibitem{Weiss} N. Weiss, Phys. Rev. D24 (1981) 475

\bibitem{Langmann} E. Langmann, M. Salmhofer and A. Kovner, 
Mod. Phys. Lett. A9 (1994) 2913 

\bibitem{Lenz2} F. Lenz, H.W.L. Naus and M. Thies, Annals Phys. 233 (1994) 317 

\bibitem{itziks} C. Itzykson, Int. Journal of Mod. Phys. A1 (1986) 65

\bibitem{bour} N. Bourbaki, Groupes et algebres de Lie, Chap 7 and 8,
Hermann, Paris, 1968; LiE sofware package, CWI, Amsterdam, 1997

\bibitem{griesshammer}H. W. Grie\ss hammer, 
\textit{Magnetic Defects Signal Failure of Abelian Projection
Gauges in QCD}, preprint FAU-TP3-97/6, hep-ph/9709462

\bibitem{thooftpolyakov} G. 't Hooft, Nucl. Phys. B79 (1974) 276; 
A.M. Polyakov, JETP lett. 20 (1974) 194

\bibitem{waia} C. Jayewardena, Helv. Phys. Acta 61 (1988) 636

\bibitem{sachs} I. Sachs and A. Wipf, Helv. Phys. Acta 65 (1992) 653;
Annals Phys. 249 (1996) 380

\bibitem{go} P. Goddard, J. Nuyts and D. Olive, Nucl. Phys. B125 (1977) 1

\bibitem{Dubrovin} B.A. Dubrovin, A.T. Fomenko, S.P. Novikov, 
\textit{Modern Geometry - Methods and Applications}, Springer Verlag New York,
1992


\bibitem{poly} A.M. Polyakov, Phys. Lett. 59B (1975) 82; 
T. Banks et.al, Nucl. Phys. B129 (1977) 493

\bibitem{seiberg} N. Seiberg and E. Witten, Nucl. Phys. B341 (1994) 484

\bibitem{lenz}
M. Engelhardt and H. Reinhardt, Phys. Lett. 430B (1998) 161;
P. van Baal and J. Keller, Annals Phys. 174 (1987) 299;
Di Giacomo, preprint IFUP-TH 9/98, hep-lat/9802008 and references therein

\end{thebibliography}
\end{document}